\newcommand{\specialcell}[2][c]{%
  \begin{tabular}[#1]{@{}l@{}}#2\end{tabular}}

\documentclass[10pt,letterpaper]{article}
\usepackage[top=0.85in,left=2.75in,footskip=0.75in]{geometry}

\usepackage{amsmath,amssymb}

\usepackage{changepage}

\usepackage[utf8x]{inputenc}

\usepackage{textcomp,marvosym}

\usepackage{cite}

\usepackage{nameref,hyperref}

\usepackage[right]{lineno}

\usepackage{microtype}
\DisableLigatures[f]{encoding = *, family = * }

\usepackage[table]{xcolor}

\usepackage{array}

\newcolumntype{+}{!{\vrule width 2pt}}

\newlength\savedwidth

\raggedright
\usepackage[aboveskip=1pt,labelfont=bf,labelsep=period,justification=raggedright,singlelinecheck=off]{caption}

\makeatletter
\renewcommand{\@biblabel}[1]{\quad#1.}
\makeatother

\date{}

\usepackage{lastpage,fancyhdr,graphicx}
\usepackage{epstopdf}
\fancyheadoffset[L]{2.25in}
\fancyfootoffset[L]{2.25in}
\usepackage{booktabs}

\begin{document}
\vspace*{0.2in}

\begin{flushleft}
{\Large
\textbf\newline{Automatic Structural Scene Digitalization} 
}
\newline
\\
Rui Tang\textsuperscript{1,2,*},
Yuhan Wang\textsuperscript{1},
Darren Cosker\textsuperscript{2},
Wenbin Li\textsuperscript{3,*}
\\
\bigskip
\textbf{1} Kujiale.com, Hangzhou, China
\\
\textbf{2} Centre for the Analysis of Motion, Entertainment Research and Applications, University of Bath, UK
\\
\textbf{3} Department of Computing, Imperial College London, UK
\\
\bigskip

%
%





* Corresponding authors, r.tang@bath.ac.uk, wenbin.li@imperial.ac.uk

\end{flushleft}
\section*{Abstract}
In this paper, we present an automatic system for the analysis and labeling of structural scenes, floor plan drawings in Computer-aided Design (CAD) format. The proposed system applies a fusion strategy to detect and recognize various components of CAD floor plans, such as walls, doors, windows and other ambiguous assets. Technically, a general rule-based filter parsing method is fist adopted to extract effective information from the original floor plan. Then, an image-processing based recovery method is employed to correct information extracted in the first step. Our proposed method is fully automatic and real-time. Such analysis system provides high accuracy and is also evaluated on a public website that, on average, archives more than ten thousands effective uses per day and reaches a relatively high satisfaction rate.



\section{Introduction}
CAD floor plans ~\cite{cadfloorplan} comprise a set of architectural drawings that describe the layout of various structural objects (e.g. walls, windows, doors and furniture) in a building.

In architecture and building design, floor plans contain various levels of detail and show the relationships among rooms, spaces and other architecture components for each level of a structure.

Floor plan analysis can be considered a special image analysis method that attempts to understand the structural and semantic information of a building by analysing 2D floor plans (in this work,{\lq}images{\rq} refers to both rasterised and vectorised images). By reviewing previous research, it is easy to conclude that there are various purposes to analysing a given floor plan. For example, several studies have applied floor plan analysis to the generation of 3D models ~\cite{dosch1999reconstruction}, ~\cite{lu2007automatic}, ~\cite{or2005highly}, and another study emphasised interpreting floor plans as CAD formats. In addition, other studies have attempted to detect rooms in architectural drawings ~\cite{mace2010system}, ~\cite{wessel2008room} and search massive floor plans ~\cite{weber2010scatch}.

~\cite{aoki1996prototype} is similar to ~\cite{llados1997system} in that they both proposed a method to understand hand-drawn floor plans. In addition, a previous study proposed a complete system for architectural diagram analysis ~\cite{dosch2000complete}, where basic primitives are recognised by applying numerous automated graphics-recognition processes. A method to detect rooms in architectural floor plan images has also been proposed ~\cite{mace2010system}. That method was then adopted and expanded ~\cite{ahmed2011improved} to include new processing steps, such as wall edge extraction and boundary detection.

This work presents an automatic system for analysing floor plan drawings in CAD format. The remainder of this paper is organised as follows. Work related is summarised in Section \ref{cad:related_work}. Section \ref{cad:system} provides an overview of the proposed method, including its specific processing steps. Section \ref{cad:evaluation} presents an evaluation of the proposed analysis method and discusses experimental results. Finally, Section \ref{cad:concl} concludes and offers suggestions for future work.

\section{Related work}

\label{cad:related_work}

Architectural drawings, typically in the form of floor plans, are necessary to design, describe and execute a construction project. The architectural elements on each building level are represented using standard symbols; thus, floor plans typically create a top-down orthographic projection.

Floor plans consist of various levels of detailed architectural elements. For example, construction structure drawings (CSD), which are one of the most complicated types of floor plan, portray internal steel bars, the concrete structure of columns, beams and WA walls, and pipe and ductwork layouts, which are popular with both design engineers and construction managers. Tong Lu and his research team ~\cite{lu2007automatic} introduced a system that constructs a detailed building model from computer-drawn CSDs. However, interpreting raster images of CSDs requires further research.

Although floor plans, which are widely used in architecture engineering and the construction life-cycle, can cover a building's complete layout, both hand-drawn and computer-produced floor plans may lack detailed construction information.

Another main drawback arises from the various graphical symbols used in floor plans. Figure \ref{fig:cad_related_work_fig_1} shows several common graphical symbols for walls, windows and doors. Note that not all floor plan drawings comply with specific standards. However, the overall purpose of floor plan drawings determines which and how components will be shown. Despite the fact that less-detailed floor plans can be considered legitimate input by many systems, the various symbols create a challenge when analysing and interpreting a floor plan image.


\textbf{Fig 1. Different ways to draw a wall with a window and a door. The variable graphic symbols pose challenges for automatically recognition of objects in CAD drawings. ~\cite{yin2009generating}}
\label{fig:cad_related_work_fig_1}

\subsection{General System Overview}
Figure 2 gives a general idea of how to automatically generate a 3D building model using a specific input. Existing systems can be categorised according to the type of input. CAD documents, e.g. Data Exchange Format (DXF) and AutoCAD Drawings, store information as 2D geometric elements in the first step and place architectural components into different groups before labelling components.

On the other hand, when a raster image floor plan is used as input, there are no obvious differences between graphical symbols, wall lines, dimensions, scales and leading lines. Thus, the system relies heavily on image-processing and pattern-recognition techniques for information extrusion.

Figure 3 shows the basic model extrusion step, and Figure 4 shows an ideal solution of a 3D model system, which differs slightly compared to most existing systems. Previous systems have inspired our common framework, which intends to help developers structure and compare current solutions.

Most existing systems lack generality, which means pattern recognition, is restricted to a small set of predefined symbols. Moreover, existing systems cannot exploit information from text strings, which normally contain useful information about a building{\rq}s spatial structure and topology. Also, current systems cannot position architectural elements appropriately, and the imperfect algorithms applied in several systems require user interaction in some steps. Thus, more accurate, efficient and automated algorithms are needed, especially for the symbol recognition step.

\subsection{Converting Floor Plan CAD files}
Systems that apply CAD-based floor plans focus more on 3D model extrusion rather than image processing and pattern recognition. Rick Lewis and Carlo Sequin  ~\cite{lewis1998generation} at the University of California, Berkeley introduced a system that creates 3D polygonal building models semi-automatically by grouping architectural symbols into specified layers in standard DXF files. Their system introduced a correction strategy on disjointed and overlapping edges in order to overcome geometric flaws. This system collects the topology of spaces and portals to generate proper polygon orientation. After each floor is modelled, the system stacks the floors to create a complete model. This system significantly simplifies the recognition process, which benefits designers in various applications, such as smoke propagation simulations.

Clifford So ~\cite{so1998reconstruction} and his colleagues from the Hong Kong University of Science and Technology (HKUST) considered the model conversion problem in a virtual reality context. They targeted three major tasks, i.e. wall extrusion, object mapping and ceiling and floor contraction, after observing model reconstruction via a conventional manual method. The processing time of their method is greatly reduced by incorporating automated approaches for each task in the next step, such as automatic wall polygon extrusion, generating and placing customised templates of random orientation and size and advancing front triangulation. However, their system has a significant disadvantage, i.e. the input file must contain fully established semantic information and no errors, which means the system requires manual intervention. For example, wall lines must be marked by users, architectural objects must be specified, and objects must be assigned to individual transformation matrices.

Researchers at the Massachusetts Institute of Technology (MIT) ~\cite{bmg} automated the construction of a realistic MIT campus model (Building Model Generation project (http://city.csail.mit.edu/bmg). Compared to the Berkeley system ~\cite{lewis1998generation} , a similar pipeline is employed; however, an additional process is used to automatically position and orient building models using a map for guidance.

Lu's research team at the Nanjing University of China proposed a system to construct models from computer-drawn CSDs and vectorised floor plans ~\cite{lu2007automatic} . Compared to other computer drawing formats, symbol recognition in a vector image is much more difficult because vector images contain unlabelled geometric primitives. Similar to the HKUST project ~\cite{so1998reconstruction} , this system differentiates walls from other architectural elements. First, it detects parallel line segment pairs as walls, which are then removed from the drawings. Next, the remaining primitives are recognised by detecting feature matches with predefined patterns that contain a symbol's graphical primitives and contextual information. In the recognition process, the system places patterns in order relative to their priority and checks each pattern one by one. Corresponding elements are removed from the drawing as soon as they satisfy all of a given pattern's constraints. Although this requires high-quality input, the system benefits users significantly because it focuses on structural details and is highly automated.

To automatically extract structural assets, correspondences e.g. optical flow~\cite{LME,li2017learn,chen2016dense,li2017video,li2017blur,li2016drift,li2014robust,li2013nonrigid,li2012anchor}, from any input CAD file to analysed files is introduced. Such correspondences give a hidden link from unknown visual elements to reference and further prorogate the actual properties back to the input CAD. Although correspondences based method may suffer from expensive time consumption, they show great potentials on fragile items digitalisation, VR based structural extraction~\cite{lv2017virtual,ren2016towards} and post-production~\cite{li2016roto,li2017nonrigid,li2016dense,tang2012global}.

\subsection{Image Parsing and Drawing Analysis}
This process analyses an input raster floor plan image and extracts layout information, which is referred to as a {\lq}parse process{\rq}. Referring to Yin{\rq}s survey ~\cite{yin2009generating}, the challenges in this step are explained in Table \ref{cad:table_1}.

Graphical document analysis technology is required to analyse and parse image floor plans, which includes two main steps: (1) removing noise ,such as text and annotation; and (2) graphical symbol recognition. The cleaning step focuses on removing noise and other irrelevant information to improve image quality. In the graphical symbol recognition step, the system categorises the recognised symbols by identifying certain information, including location, orientation and scale.

Compared to other graphical documents, floor plans have certain distinguishable features. For example, various line shapes (curved or straight) represent walls in floor plans. Another difference is that the architectural symbols are made up of simple geometric primitives. Typically, to handle such input, graphics recognition is integrated with vectorisation.
(Table \ref{cad:table_1})

\begin{table}[]
\caption{The challenges of image parsing and drawing analysis}
\label{cad:table_1}
\resizebox{\textwidth}{!}{%
\begin{tabular}{l l}
\textbf{Noise removal}      & \specialcell{Notation leading lines can easily be confused with wall lines. \\ The background may contain a grid or decorative pattern.} \\
\textbf{Text extraction}    & \specialcell{Textfont, size and orientation may vary. \\ Text and graphical symbols may share pixels (overlapping or touching).} \\
\textbf{Vectorisation}      & \specialcell{Most algorithms recover only lines and arcs. \\ Free-form curves are a challenge.\\ Noise affects the result significantly.\\ Vectorisation may yield poor results at junction points.} \\
\textbf{Symbol recognition} & \specialcell{Symbols may not comply with standards.\\ There may be a large pool of symbols, and the differences between two symbols may be subtle.}\\
\end{tabular}%
}
\end{table}

\subsubsection{Noise Removal}
Sampling noise introduced by digital scanning is very common when processing hand-drawn floor plans. However, floor plans are generated by a computer gradually; thus, noise has a broader definition in this context. For example, pixels without directly useful information are typically considered noise, including annotation leading lines, dimension lines, furniture and hardware symbols. On rare occasions, a decorative pattern in the background could be misidentified.

In Loria{\rq}s system ~\cite{survey_23} a morphological filter is applied as a fine line between noise and useful pixels. This method is based on the assumption that background patterns and dimension leading lines can be differentiated from useful lines because they have different thicknesses and styles.~\cite{or2005highly} makes a similar assumption, filtering input and only thick construction lines can be preserved.

\subsubsection{Text Extraction}
A perfect algorithm should be free from text font, size and orientation, and should be efficient and require little manual intervention. Geometric shapes mixed with text incur extra burden for separation and extraction tasks. Text research has been developed for several decades, and its results can be categorised into structural-based (focusing on structural differences) and pixel-based algorithms.

\subsubsection{Graphic Recognition}
The text is separated from graphics in the previous step. Graphic recognition is a process whereby pixels are organised and ordered according to the geometrical description of the building{\rq}s layout. Typically, architectural drawings comprise two primary types of information, i.e. structural information and local architectural components.

As shown in Table \ref{cad:table_1}, graphic recognition comprises vectorisation and symbol recognition. Walls are preserved as geometric poly-lines for the extrusion step because they define the building{\rq}s spatial structure. From this perspective, all systems introduce vectorisation and deal with geometric elements rather than performing symbol recognition on pixels directly.

\paragraph{Vectorisation}

This process, which is referred to as raster-to-vector conversion, transfers image pixels to geometric primitives. The most important aspects of each algorithm are efficiency, robustness and accuracy. The workflow of traditional line-drawing vectorisation involves two steps, as shown in the following table.

Note that correcting joint errors is required after each step. In most cases, vectorisation algorithms can find line segments and circular arcs; however, more complex curves remain a challenge for existing algorithms.

In Step 1, three groups of algorithms, i.e. parametric model fitting, contour tracking and skeletonisation ~\cite{hilaire2006robust} , are typically used. In parametric model fitting, Hough transform  ~\cite{duda1972use} is applied to detect lines; however, this requires significant amounts of memory and lacks universality.

Contour tracking detects the contour of white pixels (rather than black pixels) and recognises connected regions as rooms. This method can deal with simple floors; however, it cannot deal with complicated structures because it is based on the assumption that white spaces are divided by black wall lines in the image.

Thinning-based algorithms for skeletonisation attempt to search for a curve bones{\rq} medial axis by stripping boundary pixels until a one-pixel wide skeleton remains ~\cite{lam1992thinning} . Here, one disadvantage is that intersections always confuse the results. Another disadvantage is that thinning-based algorithms require significant time to process because each pixel is visited multiple times. Typical medial-axis-based algorithms include pixel tracking ~\cite{dori1999sparse} and run-graph-based algorithms  ~\cite{dori1999sparse} . Medial-axis-based algorithms treat a thick line as a solid shape and its medial axis as a skeleton.

In Step 2, point chains are segmented into sets of lines, poly-lines and circular arcs by estimating curvature or polygonal approximation to identify critical points.
Loria{\rq}s system introduces a skeletonisation technique and polygonal approximation to complete the vectorisation process~\cite{survey_23} . The CUHK system tracks the contour of black pixels rather than white pixels, which differs from contour tracking~\cite{or2005highly} .

\paragraph{Symbol Recognition}

This is the most important part of graphical document analysis, and the graphic symbol recogniser (GSR) in this process should be efficient and not limited to either context or affine transformation. Note that previous research has proved that several methods work well in specific areas and generate positive results.

GSRs can be classified as vector based (oriented toward structure) and pixel based (oriented toward statistics). Vector-based GSRs process graphical primitives, such as points, line segments, arcs and circles, in vectorised images. This approach checks primitives in groups to identify a symbol using region adjacency graphs ~\cite{ah2001architectural} , graphical-knowledge-guided reasoning ~\cite{yan2003engineering} , constraint networks ~\cite{ah2001architectural} and deformable templates ~\cite{valveny2003model} . Note that good vectorisation is expected with this approach, which is affine invariant.

Other GSRs are pixel based. Such recognisers process raster images without vectorisation. Such methods focus on the statistical features of a symbol{\rq}s pixel information. Pixel-based approaches contain plain binary images ~\cite{schurmann1996pattern} , living projections and shape contexts ~\cite{belongie2002shape} . Compared to vector-based approaches, pixel-based approaches are more accurate, even though their performance is sensitive to scaling and rotation. Su Yang improved a recognition method by merging pixel-based and vector-based approaches ~\cite{yang2005symbol} .

In Loria{\rq}s project, a network is applied to identify the features of a vectorised image{\rq}s primitives ~\cite{survey_23} . Then, segments in the vectorised floor plan are distributed throughout the network to find terminal symbols. A similar but simpler approach is employed in the CUHK system, in which a series of geometric constraints are considered symbol patterns. In this approach, raster or vector images of a floor plan can be used to improve recognition accuracy ~\cite{or2005highly} .

\section{Floor Plan Analysis System}
\label{cad:system}

The proposed automatic floor plan analysis system targets engineering uses and takes CAD format floor plans, which can be considered a set of vectorised images with unit information, as input.

Figure \ref{fig:CAD_work_flow} illustrates the basic workflow of the proposed floor plan analysis system, which is described in detail in the following. As mentioned previously, the proposed system is available online for engineering uses without any restrictions to the drawing style of a floor plan. This means that the proposed system can accept a wide variety of input data. Therefore, standardised parsing is required to normalise input data in the first step. After standardisation, all of the information provided by the floor plans is represented by the most basic elements, i.e. lines and arcs. Then, because floor plans represent structural data, such as walls, windows and doors, a general filter is applied to lines to obtain effective room structure information. Then, in consideration of excessive filtration in the filtering step, an image-processing-based retrieval method is adopted to correct the filtered result. Finally, the proposed system attempts to extract windows and doors from the floor plan.


\textbf{Fig 2. Work flow of the floor plan analysis system. Starting with putting in raw data, followed by the process of standardization, filtering and rasterising correcting, as a result, walls, windows and doors are detected}
\label{fig:CAD_work_flow}

\subsection{Data Standardization}
\label{CAD:Data Standardization}

\subsubsection{Problem Statement}

It is difficult to develop a general automatic system to recognise various types of CAD floor plans because designers and engineers draw floor plans in different ways. The variety of CAD floor plans can be summarised as follows.

(1) Different units are adopted in different CAD floor plans. Architecture designers employ different units (e.g. centimetres, inches and millimetres) according to the given project{\rq}s requirements or personal preference.

(2) Structural objects can comprise various internal forms. For example, as shown in Figure \ref{fig:cad_related_work_fig_1}, doors and windows can be drawn in different ways ~\cite{yin2009generating} , and, as shown in Figure 2.b, even in a single CAD floor plan, load bearing walls (filled polygons) differ from normal walls (parallel lines). Such variable graphic symbols pose challenges when attempting to recognise floor plans automatically.

(3) Dimensions are varied in CAD floor plans according to the intended purpose. For example, for architecture purposes, some CAD floor plans are shown in 3D space, while others are shown in 2D space.

(4)In most cases, manifold furniture or annotations are applied, which impede recognition of the primary structural components (i.e. walls, windows and doors). In most cases, manifold furniture or annotations are applied, which highly disturbed the recognition of main structural components (walls, windows and doors).

(5) Some CAD drawings may contain several floor plans in a single drawing (Figure 3). Each floor plan in such drawings is independent. Thus, the proposed system must be able to extract and separate the individual plans.

\subsubsection{Solutions}

A simple data standardisation process is employed to address the variety of input CAD floor plans. The primary purpose of this process is to normalise all architecture elements in the floor plans as lines. The process is described as follows.

a. The first step is to standardise units. By reading the unit information of the CAD floor plans, various units are converted to millimetres.

b. Regardless of the composition of structural objects (e.g. lines, solids, triangles, multi-lines, poly-lines or blocks), all such objects in the drawing are decomposed into lines, which are the most basic element in all architecture drawings. Simultaneously, arcs are converted to short and continuous lines.

c. The proposed system focuses on detecting walls, doors and windows in a 2D CAD floor plan. Therefore, we convert 3D floor plans into 2D spaces by calculating the normal of lines.

d. Furniture, which is considered a type of structural object, is decomposed into lines (refer to Step b). Regarding annotations, marker lines are converted into lines, and text elements are removed.

e. At this point, the architectural drawing now comprises lines and arcs. Since the input drawings can contain more than one floor plan, systematic clustering of lines is employed. The clustering classifies lines based on Euler distance. We define the distance between lines by searching the closest link between lines (Figure \ref{fig:Multiplans}).

f. Then, a 5000-mm threshold is applied, in order to cluster all lines to segment multiple floor plans from a single CAD file.


\textbf{Fig 3. An example of multiple floor plans in a single CAD drawing, systematic clustering is employed to classify lines based on Euler distance.}
\label{fig:Multiplans}

\subsection{A Fusion Strategy System for CAD Floor Plans Analysis}
Here, we introduce a fusion strategy system for CAD floor plan analysis. Floor plans always contain information that helps an architect express the actual layout of the structural objects, e.g. walls, doors and windows. However, during floor plan analysis, different types of objects must be interpreted at different points in time using specific strategies. Therefore, we introduce a fusion strategy system that combines a set of general filters and an image-parsing method to extract walls, windows and doors from a CAD floor plan. In the filtering stage, we attempt to identify as many correct walls in the floor plan as possible. Walls are one of the essential elements in a floor plan, and other architecture components, e.g. doors and windows, are attached to walls. Thus, we must design a set of filters to extract information about walls from the input data. Then, by rasterising the CAD floor plan, we attempt to restore any walls that were filtered excessively in the image-processing strategy. Based on the wall analysis results, the proposed system attempts to detect windows and doors. Then, an image-processing-based wall restoration system is employed. Finally, a mechanism is used to detect doors and windows based on the detected walls. These processes are discussed as follow.

\subsubsection{General Filters}
\label{CAD:GeneralFilters}

Similar to existing work \cite{dosch2000complete} and \cite{or2005highly}, in this step, the proposed system extracts walls from a floor plan by applying general filters. The filters are built on the assumption that walls are represented by parallel lines in Figure \ref{fig:original_floor_plan}. Based on this assumption, the filters search for parallel lines and define them as wall candidates. The workflow of the filters is explained in the following.


\textbf{Fig 4. Raw data as input of floor plan. Parallel lines are targeted in the process of filtering, base on the assumption that they represent walls.}
\label{fig:original_floor_plan}

\paragraph{1. Gradient Filter ($pi/12$)}

The objective of introducing this filter is to find non-vertical and non-horizontal lines, because based on the assumption that walls lay horizontally and vertically in a floor plan, walls can be targeted by applying this filter. Note that some designers may draw lines at a slight tilt, therefore, we set the threshold to $pi/12$. In Equation \ref{eq:2GradientFilter1}, $L_{raw}$ is the raw input line from the CAD floor plan, and $L_1$ represents filtered lines after applying the gradient filter. We then divide the filtered lines into two sets, the horizontal set ($H_{s1}$) and the vertical set ($V_{s1}$), as expressed by Equation \ref{eq:2GradientFilter2} .

\begin{equation}
\label{eq:2GradientFilter1}
L_1 = f_{Gradient_{filter}(pi/12)}(L_{std})
\end{equation}

\begin{equation}
\label{eq:2GradientFilter2}
(H_{s1},V_{s1}) = f_{splitHV}(L_1)
\end{equation}

Figure \ref{fig:2GradientFilter} shows the filtered result $L_1$  after gradient filter (threshold $pi/12$) is processed, while orange and blue lines represent horizontal set $H_{s1}$ and vertical set $V_{s1}$ respectively.


\textbf{Fig 5. Production of Gradient filter, orange lines and blue lines represent horizontal filtered lines and vertical set respectively.(Threshold: $pi/12$)}
\label{fig:2GradientFilter}

\paragraph{2. Length Filter (2mm)}

As mentioned in Section \ref{CAD:Data Standardization}, arcs are converted into short lines; however, this may generate many short lines. In the consideration of negative effects that brought by irrelevant short lines, hereby a length filter (Equation \ref{eq:2LengthFilter1}) is applied to eliminate interference by such short lines to address this issue. The threshold of this length filter set as 2mm in order to get the most accurate results and improve work efficiency as well.

\begin{equation}
\label{eq:2LengthFilter1}
H_{s2} = f_{length_{filter}(1mm)}(H_{s1}),   V_{s2} = f_{length_{filter}(1mm)}(V_{s1})
\end{equation}

In Figure \ref{fig:3LengthFilter2mm}, the red and blue lines represent the filtering result $H_{s2}$ and $V_{s2}$ after length filter is applied, respectively.


\textbf{Fig 6. Production of Length filter, red lines and blue lines represent filtering result after length filter is applied. (Threshold: $2mm$)}
\label{fig:3LengthFilter2mm}

\paragraph{3. Gap-Filling and Line Merging (1 mm, 1 mm, loop=5)}
\label{CAD:GapAndMerg115}

 In architecture drawings, small gaps or dislocations may be created when designers draw walls. The proposed system employs a gap-filling loop filter and merges close parallel lines to solve this problem. The gap-filling process attempts to connect close lines in $H_{s2}$ and $V_{s2}$ .

 In this process, it is key to determine whether such lines are sufficient close to each other, so the threshold is set to 1 mm. Then, the line-merging process merges lines in  $H_{s2}$ and $V_{s2}$ are in a specific distance. Similar to the previous stage, a threshold of the same value (1 mm) is employed in this process in order to merge close parallel lines.  Fig \ref{fig:4FillGapAndMerge} shows results obtained after applying the gap-filling and line-merging processes. In Equation \ref{eq:fillmerge1}, $H_{s3}$ and $V_{s3}$ are the filtered products of this process. Note that this process is prone to drift errors; however, small errors will be fixed Section \ref{CAD:ImageParsing}.

\begin{equation}
\label{eq:fillmerge1}
(H_{s3},V_{s3}) = f_{merge(1mm)}(f_{fill(1mm)}(H_{s2},V_{s2}))
\end{equation}


\textbf{Fig 7. Production of fill gap and merge lines processing. Gap filling process applies to lines that are close to each other within 1mm.}
\label{fig:4FillGapAndMerge}

 \paragraph{4. Removing Multiple Parallel Lines }

Commonly, sets of close multiple parallel lines in walls with an equal gap size represent windows (Figure \ref{fig:MPL_Problem}). This filter converts such multiple parallel lines in $H_{s3}$ and $V_{s3}$ into a pair of parallel lines. As shown in Figure \ref{fig:MPL_Problem} , if the outer bounds of such multiple parallel lines are connected to wall lines, a line-splitting filter is employed to split such long lines into segmented short lines.


\textbf{Fig 8. Multi-parallel lines with same gaps to represent windows. Applying a line split function to split long lines into segmented short lines because outer bounds of windows are connected in walls.}
\label{fig:MPL_Problem}

  \begin{equation}
\label{eq:rml}
(H_{s4},V_{s4}) = f_{RML_{filter}}(H_{s3},V_{s3})
\end{equation}

  In Equation \ref{eq:rml}, $H_{s4}$ and $V_{s4}$ are the the line-splitting filter.

  After the line-splitting filter is applied, inner lines in the multiple parallel lines structure are removed. Such structures must be detected in the floor plan to achieve this. Therefore, a multiple parallel lines structure detector for $H_{s3}$ and $V_{s3}$ s employed. For example, $H_{s3}$ ,the distance between lines is less than 300 mm and is marked as a benchmark, and lines in this range are placed into a candidate line group. Note that inner lines in the candidate group are removed if the number of multiple parallel lines is from three to six and the distance between each line is between 10 mm to 100 mm. Figure \ref{fig:5RML} shows results obtained after removing such multiple parallel lines.


\textbf{Fig 9. the result after removing multiple-parallel lines. Inner lines in the multiple parallel lines structure are removed after applying line-splitting filter.}
\label{fig:5RML}

 \paragraph{5. Length Filter (90mm)}
After removing multiple-parallel lines, the remaining parallel lines in $H_{s4}$ and $V_{s4}$ can be considered as candidate lines for the construction of walls. However, many irrelevant lines that do not contribute to walls can remain in sets $H_{s4}$ and $V_{s4}$ .

As described in the previous section, the proposed system attempts to find as many correct walls as possible, despite the fact that some walls may have been over-filtered.  The method used to restore walls is discussed in Section \ref{CAD:ImageParsing} .

Considering that main walls are typically represented by long pairs of parallel lines, a length filter with a threshold of 90 mm is employed to remove short lines that may cause interference. In Equation \ref{eq:lengthfilter2} , $H_{s5}$ and $V_{s5}$ denote the productions of the length filter. Figure \ref{fig:6LengthFilter90mm} shows results obtained after applying the length filter. Here, the length filter removes many pairs of short parallel lines that construct short walls.

  \begin{equation}
\label{eq:lengthfilter2}
(H_{s5},V_{s5}) = f_{length_{filter}(90mm)}(H_{s4},V_{s4})
\end{equation}


\textbf{Fig 10. The results after applying the length filter. Length filter is the process of removing pairs of short parallel lines (less than 90mm) that contribute to short wall.}
\label{fig:6LengthFilter90mm}

\paragraph{6. Connectivity Filter}

In the proposed method, a line connectivity filter is applied relative to wall continuity. First, line connectivity is determined. Then, lines that have no connection with any other lines are removed. Figure \ref{fig:7AdjacencyFilter} shows results obtained after applying the connectivity filter. In Equation \ref{eq:connfilter2}, $H_{s6}$ and $V_{s6}$ denote the productions of connectivity filter in vertical and horizontal direction respectively.

\begin{equation}
\label{eq:connfilter2}
(H_{s6},V_{s6}) = f_{connectivity_{filter}}(H_{s5},V_{s5})
\end{equation}


\textbf{Fig 11. Production of applying Connectivity Filter, where irrelevant lines are removed by applying the line-connectivity-filter method.}
\label{fig:7AdjacencyFilter}

\paragraph{7. Gap-Filling and Line Merging (90mm,50mm,loop = 5)}

For gaps between doors and long lines, which are generated by placing furniture against walls, we apply a gap-filling and line-merging filter similar to that discussed in Section \ref{CAD:GapAndMerg115} . In Equation \ref{eq:fillmerge2}, $H_{s7}$ and $V_{s7}$ denote the productions of the gap-filling filter and line-merging filter in vertical and horizontal direction respectively. The gap filling filter attempts to connect lines in $H_{s6}$ and $V_{s6}$ if they are sufficient close.
Because this filter is also apply to door gaps, The threshold is set to 90 mm when applying this system to door gaps; while threshold is set to 50mm when applying it to line merging. Consider the fact that walls generally are 120mm-240mm in width, setting 50mm as a benchmark will not disturb the result of wall selection.

In Fig \ref{fig:8Merge2} , $H_{s7}$ and $V_{s7}$ are represented by red and blue lines, respectively.

\begin{equation}
\label{eq:fillmerge2}
(H_{s7},V_{s7}) = f_{merge(50mm)}(f_{fill(90mm)}(H_{s6},V_{s6}))
\end{equation}


\textbf{Fig 12. Production of the second Fill Gap and Merge Line Processing. This is the process of filling gaps between doors and long lines. The red and blue lines represent Hs7 and Vs7 in Equation 4.9 respectively.}
\label{fig:8Merge2}

\paragraph{8. Detecting Candidate Pairs of Parallel Lines }

After the second gap-filling and line-merging filters are applied, we identify candidate pairs of parallel lines that contribute to walls. Here, two constraints are introduced. One constraint is that the distance between lines in $H_{s7}$ and $V_{s7}$ should be between 100 mm and 400 mm.

This constraint ensures that walls whose width is in this range are detected. The other constraint states that the overlapping length between each pair of lines should be greater than 400 mm because such lines are more likely to form walls. In Equation \ref{eq:fillmerge2} ,$H_{s8}$ and $V_{s8}$ denotes the productions of this step in vertical and horizontal direction respectively. Fig \ref{fig:DetectCandidatePL} shows them as red and blue lines, respectively.

\begin{equation}
\label{eq:fillmerge2}
(H_{s8},V_{s8}) = f_{pair}(H_{s7},V_{s7})
\end{equation}


\textbf{Fig 13. Identify pairs of parallel lines as candidates. Parallel lines in direction of vertical and horizontal are marked as red and blue respectively.}
\label{fig:DetectCandidatePL}

\paragraph{9. Generate Walls}

In this step, walls are generated from the candidate pairs of parallel lines in $H_{s8}$ and $V_{s8}$. Here, lines between 100 mm to 400 mm from the target line are found and such lines are considered wall candidates. Then, as shown in Figure \ref{fig:walls}, we generate walls from overlapping area; however, this method can generate incorrect walls. Therefore, such errors are fixed in the image-parsing stage (Section \ref{CAD:ImageParsing} ).


\textbf{Fig 14. Walls are generated from the candidate pairs of parallel lines, the overlapping areas that marked in yellow are walls. }
\label{fig:walls}

\subsubsection{Image-Parsing Wall Restoration System}
\label{CAD:ImageParsing}
With the above filters, although we attempt to extract as many correct walls as possible, the input data are difficult to standardised, which inevitably leads to problems. For example, the proposed system may generate incorrect walls or fail to detect correct walls due to excessive filtering. Thus, a wall restoration method based on image parsing of vectorised CAD floor plans is employed. An image-processing technique is not employed to recognise floor plans in CAD format ~\cite{lu2007automatic}~\cite{so1998reconstruction} if vectorized parsing method is applied. However, the proposed system integrates the filtered results with an image-parsing mechanism. Here, we first rasterise the CAD floor plan and the filtered results. Then, components in the floor plan image are extracted by applying an image component segmentation method. The wall restoration method is discussed in Section \ref{CAD:WallRestoring}.

\paragraph{Rasterization}
\label{CAD:ImageParsingRasterization}

The proposed method converts CAD format floor plans into images. Note that floor plans must be rasterised before the image-parsing method is applied to the vectorised CAD format floor plans. Thus, a raw CAD floor plan and the detected walls are rasterised as images ${I_{raw}}$ and ${I_{walls}}$ respectively, at equal image resolution (i.e. 4096x4096 pixels). Figure \ref{fig:Rasterized} demonstrated the rasterized result of raw data. It shows a rasterised result obtained from the raw input data, while Figure \ref{fig:WallMask} shows a rasterised result for walls extracted in the filter steps.


\textbf{Fig 15. Floor plan must be rasterized before applying image-parsing method. This figure shows a rasterised result of raw input data.}
\label{fig:Rasterized}


\textbf{Fig 16. Floor plan must be rasterized before applying image-parsing method. This figure shows a rasterised result for walls extracted in the filter steps.}
\label{fig:WallMask}

\paragraph{Component Segmentation}
\label{CAD:ImageParsingComponentSegmentation}

In floor plans, wall regions can always be clearly distinguished from other objects. Therefore, image segmentation is employed to classify the components in $I_{raw}$. In addition,  $I_{raw}$  is a binary image; therefore, the image segmentation task can be converted into labelling task for connected components in $I_{raw}$. A connected component in a binary image is a set of pixels that form a connected group. For example, the binary image (left part of Figure \ref{fig:conncomp1}) \cite{Mathwork1} has three connected components. The connected component labelling process identifies the connected components in the binary image and assigns a unique label to each component \cite{Mathwork1} (right part of Figure \ref{fig:conncomp1}).b. The proposed method employs an eight-connectivity-based two-pass connected component labelling method similar to \cite{haralock1991computer} . The pseudocode of this algorithm is shown in Figure \ref{fig:twopass}. The algorithm makes two passes over the image. The first pass assigns temporary labels and records equivalences, and the second pass replaces each temporary label with the smallest label of its equivalent class. The following is performed in the first pass.

1. We iterate through each element of the data by column then by row (Raster Scanning);

2. If the element is not the background;

a. Get the neighboring elements of the current element;

b. If there are no neighbors, uniquely label the current element and continue;

c. Otherwise, find the neighbor with the smallest label and assign it to the current element;

d. Store the  equivalent between neighboring labels.

1. The following is performed in the second pass. Iterate through each element of the data by column, then by row.

2. If the element is not the background, relabel it with the lowest equivalent label.

Figure \ref{fig:Components} scomponent labelling result $I_{rawComponent}$.


\textbf{Fig 17. Left: An example of connected components in a binary image, there are three connected components. Right: An example of labeling connected components in a binary image. Connected components in the binary image are identified before to be put a unique label respectively.}
\label{fig:conncomp1}


\textbf{Fig 18. Pseudocode of connected component labeling algorithm. Temporary equivalent labels are assigned in the first passes and the smallest label of its equivalent class will replace them in the second passes.}
\label{fig:twopass}


\textbf{Fig 19. After applying the two-pass algorithm over the image, component labeling result $I_{rawComponent}$ generated.}
\label{fig:Components}

\paragraph{Wall Restoring}
\label{CAD:WallRestoring}

The wall regions in a floor plan are distinct from other objects; therefore, wall information detected in the filter stage can be matched to labelled components in the original image $I_{raw}$. By tracking each component in the original image and comparing a single component to a wall mask, the component can be defined as a wall if more than 40\% of the wall region overlaps the wall mask.

However, the product of rough wall restored by wall candidate mask still have some outliers compare to the ground truth. Hence, in order to optimize the wall mask, we introduce two extra constraints.

The first filter considers the factor that walls have strong connectivity from each other. Hence, in this filter, we detect and remove unattached region, which is connecting to others. More specifically, unattached regions will not be identified as an effective wall if its area or length and width are less than specific thresholds (in our experiments, the threshold of length and width is 2000mm and 200mm, respectively), and it will be eliminated.

After that, the second constraint considers another factor that walls should construct a room. In the real drawing, furniture at corner includes sofa may generate parallel lines as well. Instead of connecting to other lines to generate independent region, these parallel lines may meet the condition of forming walls. If so, the occupancy rate of mask increased to 40\% plus, which would be marked as walls. To get around this problem, a limitation needs to be settled. If the region takes more than 50\% of the rectangle space that generated by plotting the minimum value and the maximum value on the X and Y axis, there is tiny possibility to be defined as walls, because walls normally do not take up large area in CAD drawing.  Moreover, in order to avoid to interfering the process of detecting small walls, the length and width of the rectangle space are much larger than width of an independent wall. In the view of the smallest room in the real world, the thresholds are set as 2000mm.

Figure \ref{fig:FINAL} shows the final wall extraction result obtained by the proposed system.


\textbf{Fig 20. wall restoration is the process of tracking each component in the original image and comparing a single component to a wall mask, and this figure shows the final wall extraction result obtained by the proposed system.}
\label{fig:FINAL}

\subsubsection{Detect Windows and Doors}
\label{CAD:DetectWindowsDoors}
\paragraph{Door Detection}

Typically, doors are attached to walls, i.e. they do not exist independently. Thus, doors are identified by detecting arcs that are close to detected walls. Normally, arcs with a 90 degree central angle whose radius is 300 mm to 25000 mm are considered doors.

\paragraph{Window Detection}

Window detection is similar to door detection because windows and doors are both attached to walls. In the proposed method, windows are detected as a part of a wall (Section \ref{CAD:GeneralFilters}). Thus, it is possible to find windows by searching multiple parallel lines that are less than 20 mm apart. Also, if the distance between the central line of a group of multiple parallel lines and the central line of its corresponding wall is less than one-quarter of the wall thickness, this group of multiple parallel lines is considered a window.

\section{Evaluation}
\label{cad:evaluation}

It is hard to quantitatively evaluate the CAD extraction system because there is barely proper ground truth publicly available. To give an intuitive sense on precision of our system, we implemented the proposed architectural drawing recognition system using Java; we then deploy the proposed system onto Kujiale.com which is a Chinese leading Internet company in interior design. In this case, The proposed system, which is freely available online, incorporates a mature human-computer interaction and representation system. Users can obtain recognition results by uploading CAD floor plans (DWT or DWG format). Figure \ref{fig:results} shows samples from real customers who actually perform analyzing CAD floor plans for three different real-life projects.  Please note that such components extraction is achieved without user intervention.


\textbf{Fig 21.\label{fig:results} The result of analyzing CAD floor plans for three different real-life projects. The evaluation result proves that the system is able to complete recognition process without user intervention.}

To evaluate the efficiency of our system, we record the usage of every system call then obtain the time consumption (on a mid-ranked Xeon server). On average, the proposed system requires only 2.3 second to extract information from a CAD floor plan. Compared to Lu's \cite{lu2007automatic} system, which requires nearly one hour to parse a document with 72000 graphic primitives, the proposed system requires approximately 5 seconds to parse a complicated floor plan with a massive number of objects, e.g. more than 35000 lines. The improved efficiency of the proposed system is primarily due to the high-performance algorithm.

Moreover, as mentioned previously, the proposed system can detect a large proportion of walls in most cases. Because the proposed system is intended for practical application by different types of designers, it is difficult to obtain a ground truth for all floor plans. However, due to the lack of a CAD floor plan database, the proposed system was evaluated in a user study. We asked users to score the recognition results of our system (1 means not satisfied and 10 means very satisfied). As shown in Figure \ref{fig:feedback4}, based on the research, we get an average score at 7.71 from 2515 user study samples, which indicates an outstanding performance of the system

Meanwhile, according to the statistic on CAD recognition system, the number of recognition requests fluctuated at 80,000 each week between March and July in 2017. The figure bottomed at 70,000 on week of 4th April and reached the peak at approximately 98,000 on week of 13th June.


\textbf{Fig 22. Based on the research, we get an average score at 7.71 from 2515 user study samples, which indicates an outstanding performance of the system}
\label{fig:feedback4}


\textbf{Fig 23. Statistic of our CAD recognition System. There are over ten thousands request per day.}
\label{fig:feedback1}

\section{Conclusion and Further Work}
\label{cad:concl}

A complete system for automatic detection and labeling of architectural elements from CAD format floor plan drawings has been proposed. The proposed system consists of structural and semantic analysis processes to identify relevant information and filter irrelevant information simultaneously. By applying these processes, useful architectural components, such as walls, windows and doors, can be extracted.

Although the theory behind such systems has frequently been discussed over the last two decades, no mature system has been developed for general use. Before the release of the proposed system, manual processes have dominated the market (AutoCAD Revit~\cite{revit} and Chief Architect~\cite{chiefarchitect}). However, the results of a user evaluation (4500 uses per day) demonstrate that the proposed system is efficient and effective.

In China, the large population and increasing urbanisation have increased the need for new housing, which has stimulated rapid real estate development. In consideration of the proposed system{\rq}s contributions, we believe that it can have significant economic influence by benefiting the architectural industry.

The discussions and evaluations presented in this paper have demonstrated that the proposed system outperforms an existing method by reducing processing time to around 2 or 3 seconds. In the user study, the proposed system achieves an impressive satisfaction rate (90\%).

Based on previous discussion and evaluation, our proposed system outperforms previous method by reducing processing time to just 2 or 3 seconds. Meanwhile, the system achieves an impressive satisfaction rate that nine out of ten of the targeted users are satisfied with the result.

In future, we plan to improve the proposed system in several ways. For example, currently, the proposed system is weak when detecting arc walls. Thus, further research into detecting arcs more efficiently is planned. Another problem is that, if the user defines the unit in a floor plan incorrectly, the proposed system will fail to extract anything from the floor plan. Therefore, this issue must be addressed in future. Furthermore, constructing 3D models from 2D floor plans is required by both designers and scientists; thus, we also plan to address the 3D model construction issue.

{\small
\bibliographystyle{acm}
\bibliography{documents}
}

\end{document}